\documentstyle[12pt]{article}

\newcommand{\be}{\begin{equation}}
\newcommand{\ee}{\end{equation}}
\newcommand{\ba}{\begin{eqnarray}}
\newcommand{\ea}{\end{eqnarray}}
\newcommand{\la}{\lambda}

\newcommand{\tr}{\rm tr}

\begin{document}
%\hoffset=-.4truein\voffset=-0.5truein
%\setlength{\baselineskip}{14pt}
\setlength{\textheight}{8.5 in}
\begin{titlepage}
\begin{center}
\hfill LPTENS 08-25\\
\vskip 0.6 in
{\large \bf  { Intersection numbers from the antisymmetric Gaussian matrix model
}}
\vskip .6 in
\begin{center}
{\bf Edouard Br\'ezin$^{a)}$}{\it and} {\bf Shinobu Hikami$^{b)}$}
\end{center}
\vskip 5mm
\begin{center}
{$^{a)}$ Laboratoire de Physique
Th\'eorique, Ecole Normale Sup\'erieure}\\ {24 rue Lhomond 75231, Paris
Cedex
05, France. e-mail: brezin@lpt.ens.fr{\footnote{\it
Unit\'e Mixte de Recherche 8549 du Centre National de la
Recherche Scientifique et de l'\'Ecole Normale Sup\'erieure.
} }}\\
{$^{b)}$ Department of Basic Sciences,
} {University of Tokyo,
Meguro-ku, Komaba, Tokyo 153, Japan. e-mail:hikami@dice.c.u-tokyo.ac.jp}\\
\end{center}     
\vskip 3mm         

{\bf Abstract}                  
\end{center}
The matrix model of topological field theory for the moduli space of p-th spin 
curves is extended to the case of the Lie algebra of the orthogonal group. We derive a new duality relation for the expectation values of characteristic polynomials in the  antisymmetric Gaussian
matrix model with an external matrix source.
The intersection numbers for  non-orientable surfaces of spin curves with k marked points are obtained
from the Fourier transform of the k-point correlation functions at the critical point where
the gap is closing.

\end{titlepage}
\vskip 3mm

%*******************************
\section{Introduction}

For  Riemann surfaces, it is known from Witten's conjectures \cite {Witten} and Kontsevich's derivation \cite{Kontsevich} that  the intersection numbers  of the moduli space of curves with marked points
may be   obtained from an 
Airy matrix model. Furthermore,
 higher Airy matrix models have been shown to give 
the intersection numbers of the moduli space for p-spin curves \cite{Witten1}.

Recently, a duality relation has been applied to this problem
\cite{BH1,BH2,BH3}. The derivation relied on a duality between the higher p-th Airy matrix models and  Gaussian matrix models in an external matrix source at 
critical values of this source \cite{BH7,BH8} ;  
the intersection numbers are then easily obtained from this dual model. When p=2, the model reduces to 
Kontsevich's model ; its dual connects to the behavior of  correlation functions 
near the edge of the semi-circle spectrum \cite{Okounkov,BH1}.

The  moduli space of p-th spin curves is described by random  Hermitian matrices, the
Lie algebra of the unitary group $U(N)$.
It is of  interest to extend this moduli space of spin curves for  non-orientable 
surfaces, both in the fields of  open string theory and  of  quantum chaos.

For  obtaining non-orientable
surfaces from standard loop-expansions of matrix models,  a first possibility would be to use real symmetric matrices. 
The Euler characteristics of the moduli spaces of real algebraic curves with  marked points 
may be obtained from the real symmetric matrix model \cite{Goulden}. 
However, for the intersection numbers of 
the moduli space of curves with marked points, 
this real symmetric matrix model remains difficult to solve  when one 
extends Kontsevich Airy matrix model to  non-orientable surfaces \cite{Gross}.

In this article we have chosen, instead of  real symmetric matrices, 
 to consider real antisymmetric matrices, the Lie algebra
of the $SO(N)$ group (we assume that $N$ is an even integer). Let us note that
the Gaussian random matrix model 
of the classical groups $O(N)$ and $Sp(N)$ appeared earlier in the literature
in the studies of the moments of the ${\mathcal L}$-functions \cite{BH4}
and in the study of the spectrum of  excitations inside  superconducting vortices 
\cite{BHL}.

For the $O(N)$ matrices, there is a  HarishChandra formula for the integrals 
over  the orthogonal group. Thanks  to this integral formula, which is similar to the unitary case, 
generating functions of  the intersection numbers for non-orientable surfaces  become calculable.
We shall discuss a duality relation for the $O(N)$ case, which is surprizingly similar to the $U(N)$ case ; 
then we compute  explicit expansions for the Fourier transforms of the correlation
functions of the dual models, and obtain the intersection numbers.
This study may shed  some light on the moduli space of curves on non-orientable surfaces.

\vskip 5mm
\section{Duality relation}
\setcounter{equation}{0}
\renewcommand{\theequation}{2.\arabic{equation}}

Let us first state the basic duality relation which will be used in this article. 
{\bf Theorem 1:}
\be\label{eq1}
< \prod_{\alpha=1}^k {\rm det}(\lambda_\alpha\cdot {\rm I} - X )>_A
= <\prod_{n=1}^N {\rm det}( a_n\cdot {\rm I} - Y ) >_\Lambda
\ee
where $X$ is $2N\times 2N$ real antisymmetric matrix ($X^{t} = - X$ ) 
and $Y$ is $2k\times 2k$ real 
antisymmetric matrix ;  the eigenvalues of $X$ and $Y$ are thus pure imaginary. $A$ is also a $2N \times 2N$ antisymmetric matrix, and it couples to
X as an external matrix source.  The matrix $\Lambda$ is $2k\times 2k $ antisymmetric matrix, coupled to $Y$.
We assume, without loss of  generality, that $A$ and $\Lambda$ have the  canonical form : 

\be\label{canonical}
A = \left( \matrix{ 
0 & a_1 & 0 & 0&\cdots\cr
-a_1
&0& 0& 0&\cdots\cr
0&0&0& a_2&0\cr
0&0&-a_2&0&0\cr
\cdots} \right),
\ee
i.e.
\be
A = a_1 v \oplus  \cdots \oplus a_N v, \hskip 5mm v =i\sigma_2 = \left( \matrix{0&1\cr -1&0}\right ).
\ee
$\Lambda$ is expressed also as
\be
\Lambda = \lambda_1 v \oplus \cdots \oplus \lambda_k v.
\ee
The characteristic polynomial $ {\rm det}(\lambda\cdot {\rm I}-X)$ has the $2N$ roots 
$(\pm i\la_1, \cdots, \pm i\la_n)$. 
The Gaussian averages in  (\ref{eq1}) are defined as 
\ba
< \cdots>_A &=& \frac {1}{Z_A}\int dX e^{\frac{1}{2}{\tr} X^2 + {\tr X A}}\nonumber\\
<\cdots>_\Lambda &=&\frac{1}{Z_{\Lambda}} \int dY e^{\frac{1}{2}{\tr}Y^2 + {\tr Y \Lambda}}
\ea
in which $X$ is a $2N\times2N$ real antisymmetric matrix, and $Y$ a $2k\times2k$ real antisymmetric matrix ; the coefficients $Z_A$ and $Z_{\Lambda}$ are such that the expectation values of one is equal to one. The derivation of Theorem 1 relies on a representation of the characteristic polynomials in terms of integrals over Grassmann variables,  as  for the  $U(N)$  
   or $U(N)/O(N)$ \cite{BH4,BH5,BH6} cases. Given the complexity of 
the intermediate steps of the derivation for the $O(N)$ case, the simplicity of
 the result is striking. The derivation is given in   appendix A. 
%%%%%%%%%%%%%%%%%%%%%%%%%%%%%%%%%%%%%%%%%%%%%%%%%%%%%%%%%%%%%%%%%%%%%%%%%
\section{ Higher Airy matrix models}
\setcounter{equation}{0}
\renewcommand{\theequation}{3.\arabic{equation}}

From  the theorem 1, we can obtain easily the  higher Airy matrix models. We first consider the simple case in which  the source A is a multiple of identity : $a_n=a$. Then  we write
\ba
< \prod_{i=1}^N {\rm det} (a_n\cdot {\rm I} - Y) >_\Lambda &=& 
< [{\rm det} (a\cdot {\rm I} - Y)]^N >_\Lambda \nonumber\\
&=& \frac{1}{Z_\Lambda}\int dY e^{N {\tr}{\rm log}( a\cdot {\rm I}- Y) + \frac{1}{2}{\tr}Y^2 + {\tr}Y \Lambda}
\ea
Expanding the logarithmic term, and noting that 
the traces of odd powers of $Y$ vanish 
since $Y$ is antisymmetric, 
we obtain
\be
< [{\rm det} (a - Y)]^N >_\Lambda
= \frac{1}{Z_\Lambda}\int dY e^{ 2 k N {\rm log} a - (\frac{N}{2 a^2}-\frac{1}{2}){\tr} Y^2-\frac{N}{4a^4}{\tr} Y^4  
+ \cdots +{\tr}Y \Lambda}
\ee
Chosing $ a^2 =  N$, the coefficient of ${\tr} Y^2$ vanishes. 
Then one rescales  $Y \to N^{\frac{1}{4}} Y$, and  $\Lambda \to N^{-\frac{1}{4}}\Lambda$.
After these rescalings, we obtain in the large N limit,
the higher Airy matrix model,
\be\label{Z_3}
Z = \int dY e^{-\frac{1}{4}{\tr} Y^4 + {\tr} Y\Lambda}
\ee
Note that higher powers of $Y^{2n}$ disappear in this scaling limit since they are given by 
\be
\frac{1}{n N^{n-1}}\cdot N^{\frac{2n}{4}} {\tr} Y^{2n} \sim
N^{-\frac{n}{2}+1} {\tr}Y^{2n}
\ee
which vanish in the large-N limit for $n>2$. 

By appropiate tuning of the  $a_n$'s, and corresponding  rescaling of $Y$ and $\Lambda$, 
one may generate similarly  higher models of type 
\be\label{Z_p}
Z= \int dY e^{-\frac{1}{p+1}{\tr} Y^{p+1} + {\tr} Y \Lambda}
\ee
where $p$ is an odd integer.
These models are similar to the generalized Kontsevich model in the unitary case, which gives
the intersection numbers of the moduli spaces of p-th spin curves.
However, the matrix $Y$ being real and antisymmetric, the partition function $Z$ is
very different from the unitary case and non-orientable surfaces lead to different intersection numbers.

\section{ Expansion in inverse powers of Lambda}
\setcounter{equation}{0}
\renewcommand{\theequation}{4.\arabic{equation}}

The free energy $F= {\rm log} Z$ can be expanded in powers 
of ${\tr}\Lambda^{-m}$ as in the unitary case. This is done through the
Harish Chandra formula \cite{HC} for the integration over the orthogonal group
$g= SO(2N)$.
We may take $Y$ and $\Lambda$ in  canonical form  (\ref{eq1}) without loss of generality 
: then the Harish Chandra integral reads \cite{HC}

{\bf Theorem 2 (Harish Chandra) :}
\be\label{Th2}
\int_{SO(2N)} dg e^{{\tr}( g Y g^{-1} \Lambda)} = C
\frac{\sum\limits_{w \in W} ({\rm det}w){\rm exp}[2 \sum\limits_{j=1}^N w(y_j)\lambda_j]}{
\prod\limits_{1\leq j<k \leq N}(y_j^2-y_k^2)(\lambda_j^2-\lambda_k^2)}
\ee
where  $C=(2N-1)!\prod_{j=1}^{2N-1}(2j-1)!$, and $W$ is  the Weyl group,
which consists here of permutations  followed  by reflections ($y_i\to \pm y_i\hskip 2mm ; i=1,\cdots ,N$) with an even number of sign changes.

In the appendix B, we give a more  explicit determinantal expression  for  this HarishChandra 
integral  for the orthogonal group $SO(2N)$. 

For the $p=3$ critical model (\ref{Z_p}), 
let us compute the perturbation expansion of the free energy $F= \rm{log} Z$. 
From Th. 2, we obtain (see appendix B)
\be \label{postHC}
Z = \int \prod_{i=1}^k dy_i \frac{\prod(y_i^2-y_j^2)}{\prod (\lambda_i^2-\lambda_j^2)}
e^{-\frac{1}{2}\sum y_i^4 - 2\sum y_i \lambda_i}
\ee
To obtain the series in powers of $1/\lambda$, we change $\lambda_i\rightarrow \lambda_i^3$ and
make a shift $y_i\rightarrow y_i-\lambda_i$ to eliminate the $y_i\lambda_i^3$ term. Then, the problem
reduces to cubic and quartic perturbations with a  Gaussian weight.
For instance, $k=2$ for $O(2k)$ case, we have from Th.2,
\ba
&&\int dy_1dy_2 \frac{y_1^2-y_2^2}{\lambda_1^2-\lambda_2^2} e^{-\frac{1}{2}(y_1^4+y_2^4)- 2
(y_1\lambda_1^3+y_2\lambda_2^3)}\nonumber\\
&&= 1+ \frac{1}{72}(\frac{1}{\lambda_1^4} + \frac{1}{\lambda_2^4})+ 
\frac{1}{12}(\frac{1}{\lambda_1^2}
+\frac{1}{\lambda_2^2})^2 + O(\frac{1}{\lambda^8})
\ea
where we have dropped a normalization constant.  In that simple $k=2$-case 
it is easy to make this calculation directly without
use of Th.2, and the results do agree. The coefficients $\frac{1}{72}$ and 
$\frac{1}{12}$ are universal
factors independent of $k$. For  general $k$, it is useful to write the 
Vandermonde product as a determinant
with the  $y_i^2$, and the evaluation is straightforward.
We introduce the following parameters, similar to those of  the unitary case,
\be\label{first}
t_{n,j} = (p)^{\frac{j-p-n(p+2)}{2(p+1)}}
\prod_{l=0}^{n-1}(lp+j+1 )\sum_{i=1}^k \frac{1}{\lambda_i^{pn+ j + 1}}
\ee
with here p=3. Note that the  normalization of the  $t_{n,j}$ is 
slightly different from the unitary case, in which $p$ is replaced by $- p$ in the first factor of 
(\ref{first}). However, this definition is not approriate for the half-integer genus (non orientable surface),
since the first factor becomes irrational number. So, we define in this half-integer genus  case (condition
will appear below in (\ref{g})) as
\be\label{half}
t_{n,j} = \prod_{l=0}^{n-1}(lp+j+1 )\sum_{i=1}^k \frac{1}{\lambda_i^{pn+ j + 1}}.
\ee
The index  $n_i$ stands for the power of the first Chern class $c_1$.
The index $j$ is the spin index, which takes the values $j=1,2,...,p-1$.

We end up with an expansion,   similar to  the unitary case, but with different coefficients
\be\label{intersection}
{\rm log} Z = \sum < \prod \tau_{n_i,j_i}^{d_{n_i,j_i}} > \prod \frac{t_{n_i,j_i}^{d_{n_i,j_i}}}{d_{n_i,j_i}!}
\ee

For $p=3$, the lowest orders of the $O(N)$ model are given by 
\ba\label{logZ}
{\rm log} Z &=& \frac{1}{72} \sum \frac{1}{\lambda_i^4}+ \frac{1}{12}(\sum \frac{1}{\lambda_i^2})^2
+ \frac{5}{432}\sum \frac{1}{\lambda_i^8}+ 
\frac{1}{432}(\sum \frac{1}{\lambda_i^4})^2\nonumber\\
&&+ \frac{1}{36}(\sum \frac{1}{\lambda_i^4})(\sum \frac{1}{\lambda_i^2})^2
-\frac{1}{108}(\sum \frac{1}{\lambda_i^2})^4
+ O(\frac{1}{\lambda^{12}})
\ea
Note that there is no  odd-power of $1/\lambda$ such as $\sum \frac{1}{\lambda_i}$. 
This is due to  the parity $\lambda_i\rightarrow -\lambda_i$ for real antisymmetrix matrices of $O(2N)$.
From the above series, the intersection numbers $<\prod_{n_i,j_i} \tau_{n_i,j_i}^{d_{n_i,j_i}}>$ are obtained.
In the unitary case, they are  given by
\be
<\tau_{n_1,j_1}\cdots \tau_{n_s,j_s}> =
\frac{1}{p^g}\int_{\bar M_{g,s}} \prod_{i=1}^s c_1({\mathcal L}_i)^{n_i}
C_T(j_1,\cdots,j_s)
\ee
where $c_1$ is the first Chern class and $C_T(V)$ is the top Chern class \cite{Witten1}.
In the present case, we call the intersection numbers as the coefficients of the expansion of ${\rm log}Z$ as
(\ref{intersection}).

The numbers of $\tau$ corresponds to the numbers of marked points $s$. The indices $n_i,j_i$ are related to 
the genus $g$.
\be\label{g}
\sum_{i=1}^s (n_i + \frac{1}{p}j_i -1) = (3- (1- \frac{2}{p}))(g-1)
\ee
The genus $g$ is given through an expansion in powers of  the inverse of the size of the matrix, as is standard  for matrix models. 
For this purpose, we introduce an overall  factor $k$ in the exponent (\ref{Z_p}), an integral over $k\times k$ matrices. 
Then ${\rm log}Z/k$ may be expanded in a series in powers of  $k^{2-2g}$, with genus $g$. For the present antisymmetric
matrix $Y$,  odd powers of $1/k$ are also present. (In the unitary case, only even powers appear).

The genus $g$ is given by the Euler characteristics, 
\be
V - E + F = 2 - 2 ({\rm type})
\ee
where $V$, $E$ and $F$ are the numbers of vertices, edges and faces, respectively.
For  non-orientable surfaces, the genus  $g$ is replaced by the (type) : we can still use $g$ 
but it takes half-integer values. This definition coincides with that of (\ref{g}).
In Fig.1-Fig.4, the lower order terms $<\tau_{1,0}>_{g=1}$((1)torus and (2)Klein bottle),
$<\tau_{0,1}^2>_{g=\frac{1}{2}}$ ((3)projective plane), and   $<\tau_{2,1}>_{g=\frac{3}{2}}$
( crosscapped torus) are depicted.

For non-orientable surface , new characteristic terms are present, such as 
$t_{1,0}$ for the  Klein bottle, 
in addition to the torus, and  $t_{0,1}^2$ (projective plane), $t_{2,1}$ (crosscapped torus)
 which did not exist  in the
unitary case.
From (\ref{logZ}), we have the intersection numbers,
\be
< \tau_{1,0} >_{g=1} = \frac{1}{24}, \hskip 2mm <\tau_{0,1}^2 >_{g=\frac{1}{2}} = \frac{1}{6},\hskip 2mm
 <\tau_{2,1} >_{g=\frac{3}{2}} = \frac{1}{864}, \hskip 2mm <\tau_{1,0}^2 >_{g=0} = \frac{1}{24}.
\ee

\begin{picture}(100,230)(10,20)
\put (20.0,100.0){\line(1,0){100.0}}
\put (20.0,20.0){\line(1,0){100.0}}
\put (20.0,20.0){\line(0,1){80.0}}
\put (120.0,20.0){\line(0,1){80.0}}
\put (70.0,17.0){$>$}
\put (75.0,17.0){$>$}
\put (50.0,5.0){(1)torus}
\put (70.0,97.0){$>$}
\put (75.0,97.0){$>$}
\put (16.0,50.0){$\wedge$}
\put (116.0,50.0){$\wedge$}
\put (20.0,60.0){\dashbox{5.0}(45,40)[t]}
\put (70.0,60.0){\dashbox{5.0}(50,40)[t]}
\put (20.0,20.0){\dashbox{5.0}(45,35)[t]}
\put (70.0,20.0){\dashbox{5.0}(50,35)[t]}
%%%%%%%%%%%%%%%%%%%%%%
\put (140.0,100.0){\line(1,0){100.0}}
\put (140.0,20.0){\line(1,0){100.0}}
\put (140.0,20.0){\line(0,1){80.0}}
\put (240.0,20.0){\line(0,1){80.0}}
\put (190.0,17.0){$>$}
\put (195.0,17.0){$>$}
\put (170.0,5.0){(2)Klein}
\put (190.0,97.0){$>$}
\put (195.0,97.0){$>$}
\put (136.0,50.0){$\wedge$}
\put (236.0,50.0){$\vee$}
\put (140.0,60.0){\dashbox{5.0}(45,40)[t]}
\put (190.0,60.0){\dashbox{5.0}(50,40)[t]}
\put (140.0,20.0){\dashbox{5.0}(45,35)[t]}
\put (190.0,20.0){\dashbox{5.0}(50,35)[t]}
%%%%%%%%%%%%%%%%%%%%%%
\put (260.0,100.0){\line(1,0){100.0}}
\put (260.0,20.0){\line(1,0){100.0}}
\put (260.0,20.0){\line(0,1){80.0}}
\put (360.0,20.0){\line(0,1){80.0}}
\put (310.0,17.0){$>$}
\put (315.0,17.0){$>$}
\put (260.0,5.0){(3)projective plane}
\put (310.0,97.0){$<$}
\put (315.0,97.0){$<$}
\put (256.0,50.0){$\wedge$}
\put (356.0,50.0){$\vee$}
\put (260.0,60.0){\dashbox{5.0}(45,40)[t]}
\put (310.0,60.0){\dashbox{5.0}(50,40)[t]}
\put (260.0,20.0){\dashbox{5.0}(45,35)[t]}
\put (310.0,20.0){\dashbox{5.0}(50,35)[t]}
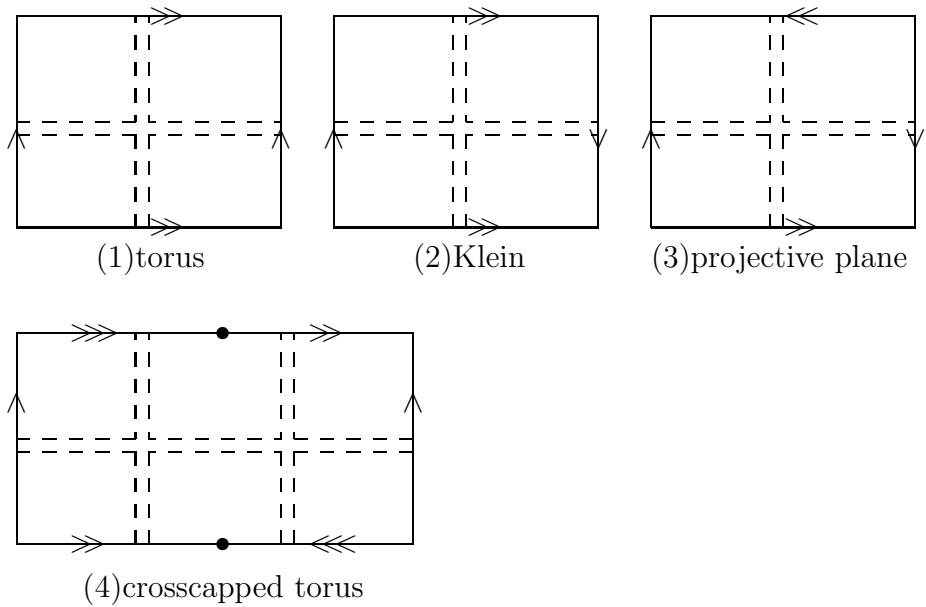
\put (20.0,130.0){{\bf{Fig.1-Fig.4:}} opposite lines  are glued together with matching arrow}
%%%%%%%%%%%%%%%%%%%%%%%%%%%
\put (20.0,-20.0){\line(1,0){150.0}}
\put (20.0,-100.0){\line(1,0){150.0}}
\put (20.0,-100.0){\line(0,1){80.0}}
\put (170.0,-100.0){\line(0,1){80.0}}
\put (40.0,-103.0){$>$}
\put (45.0,-103.0){$>$}
\put (45.0,-120.0){(4)crosscapped torus}
\put (40.0,-23.0){$>$}
\put (45.0,-23.0){$>$}
\put (50.0,-23.0){$>$}
\put (130.0,-23.0){$>$}
\put (135.0,-23.0){$>$}
\put (130.0,-103.0){$<$}
\put (135.0,-103.0){$<$}
\put (140.0,-103.0){$<$}
\put (16.0,-50.0){$\wedge$}
\put (166.0,-50.0){$\wedge$}
\put (95.0,-23.0){$\bullet$}
\put (95.0,-103.0){$\bullet$}
\put (20.0,-60.0){\dashbox{5.0}(45,40)[t]}
\put (70.0,-60.0){\dashbox{5.0}(50,40)[t]}
\put (20.0,-100.0){\dashbox{5.0}(45,35)[t]}
\put (70.0,-100.0){\dashbox{5.0}(50,35)[t]}
\put (125.0,-100.0){\dashbox{5.0}(45,35)[t]}
\put (125.0,-60.0){\dashbox{5.0}(45,40)[t]}

\end{picture}
\vskip 20mm
%%%%%%%%%%%%%%%%%%%%%%%%%%%%%%%%%%%%%%%%%%%%%%%%%%%%%%%%%%%%%%%%%
\vskip 35mm
\section{Evolution operators at  edge singularities}
\setcounter{equation}{0}
\renewcommand{\theequation}{5.\arabic{equation}}

We have derived the higher Airy matrix model of (\ref{Z_p}) from the large N limit 
of the characteristic polynomials for the
antisymmetric matrix $Y$, which is dual to the characteristic polynomial of $X$.
The choice of the $a_n = a_c = \sqrt{N} $ for (\ref{Z_3}) corresponds to  
a singular point  in the spectrum 
of  eigenvalues of $X$. At that critical point the density of states of 
$X$ has a singularity at the origin. For $a>a_c$ there is a gap
at the origin in the spectrum (whose support lies on the imaginary axis), and at $a=a_c$ 
this gap is closing. This happens also in  the unitary case for an external 
matrix source with eigenvalues $\pm a$ \cite{BH7,BH8}.

An  integral representation for the evolution operators $U$ for the vertices of O(N) (N even) may be obtained from the
Fourier transform of the correlation functions. The derivations are given in appendix C.
The evolution operators $U(s_1,...,s_n)$ are  defined as 
\be
U(s_1,...,s_n) = \frac{1}{N}< {\tr} e^{s_1 X} {\tr}e^{s_2 X} \cdots {\tr} e^{s_n X} >_A
\ee
Fot the one point function of $(2N)\times (2N)$ antisymmetric matrix $X$, we have from (\ref{C1}),
\be\label{U(s)}
U(s) =  \frac{1}{2N}<{\tr} e^{s X}>_A=- \frac{1}{Ns} \oint \frac{dv}{2\pi i}
\prod_{i=1}^N\frac{v^2 + a_i^2}{(v+ \frac{s}{2})^2 + a_i^2}
\left(\frac{v+\frac{s}{2}}{v+\frac{s}{4}} \right)e^{sv+\frac{s^2}{4}}
\ee 

For a  $N\times N$ real antisymmetric matrix $X$, the sourceless probability density
$A=0$, 
\be
P(X)=\frac{1}{Z}e^{\gamma {\tr}X^2},\hskip 5mm Z = \left(\frac{\pi}{2\gamma}\right)^{\frac{N(N-1)}{4}}
\ee
gives the  expectation values 
\ba\label{trX}
&&<X_{ij}X_{kl}> = -\frac{1}{4\gamma}(\delta_{ik}\delta_{jl}-\delta_{il}\delta_{jk})\nonumber\\
&&<{\tr} X^2 > = -\frac{N(N-1)}{4\gamma}\nonumber\\
&&<({\tr}X^2)^2> = \frac{N(N-1)(N^2-N+4)}{16\gamma^2}\nonumber\\
&&< {\tr} X^4 > = \frac{N(N-1)(2N-1)}{16\gamma^2}
\ea
We have thus to compare
\be\label{UN} 
U(s) = \frac{1}{N}<{\tr}e^{sX}>=1 + \frac{s^2}{2N}<{\tr}X^2> + \frac{s^4}{4!N}<{\tr}X^4> + \cdots \nonumber
\ee
with this integral representation. For instance, in the case of N=1, (and  $\gamma =1/2$)
the formula (\ref{U(s)}), after taking the residue at $v= -\frac{s}{2}$, leads to 
\be
U(s) = e^{-\frac{s^2}{4}}= 1 - \frac{s^2}{4} + \frac{s^4}{32}+\cdots
\ee
which indeed agrees with $N=2$ in (\ref{trX}),
\ba
\frac{1}{N}<{\tr}X^2 >  &= &-\frac{(N-1)}{2}{\Big\vert}_{N=2} = -\frac{1}{2}, \nonumber \\
\frac{1}{N}<{\tr}X^4 >  &= &\frac{(N-1)(2N-1)}{4}{\Big\vert}_{N=2} =\frac{3}{4}
\ea
For $N=2,3,...$, it is easily verified
 that the integral representation of $U(s)$ agrees with (\ref{trX}).

In a previous article, we have found an explicit formula giving the zero-replica 
limit $N\rightarrow 0$ for $U(s_1,...,s_n)$ in the unitary case, in the absence of 
an external source ($A$ =0)
\cite{BH2}. There it was shown that 
\be
\lim\limits_{N\to0} U(s_1,...,s_n) = \frac{1}{\sigma^2}
 \prod_{i=1}^n 2 {\rm sh} \frac{s_i \sigma}{2}
\ee
where $\sigma = \sum\limits_{i=1}^n s_i$.
For n=1, this is simply 
\be
\lim\limits_{N\rightarrow 0} U(s) = \frac{2}{s^2}{\rm sh}\frac{s^2}{2} = 1 + \frac{s^4}{24} + \frac{s^8}{5!\cdot 2^4}+ \cdots
\ee
This means that
\be
\lim\limits_{N\rightarrow 0} \frac{1}{N}<{\tr} M^4> = 1.
\ee
The replica limit counts the numbers of diagrams which can be drawn in one stroke line, 
and it corresponds to one marked point
for the intersection numbers.

For  real antisymmetric matrices $X$, this replica limit of $U(s)$ is different from the previous  
unitary case.
From (\ref{U(s)}), we obtain
\ba\label{N0U}
\lim\limits_{N\rightarrow 0} U(s) &=& \frac{2}{s}\oint \frac{dv}{2\pi i}{\rm log}(1 + \frac{s}{2v})
\left(\frac{v+\frac{s}{2}}{v+\frac{s}{4}}\right) e^{s v + \frac{s^2}{4}}\nonumber\\
 &=&\frac{4}{s^2}{\rm sh} \frac{s^2}{4} + 
\int_0^{\frac{s^2}{4}} dx \frac{{\rm sh}x}{x}
\nonumber\\
&=& 1 + \frac{s^2}{4} + \frac{s^4}{96} + \frac{s^6}{1152}+\cdots
\ea
The coefficients of $s^2$ and $s^4$, $\frac{1}{4}$ and $\frac{1}{96}$, are consistent
with (\ref{UN}) by (\ref{trX}), when we take $\gamma=\frac{1}{2}$ and $N\rightarrow 0$.
The term of order $s^2$ is a M\"obius band (projective plane), 
and it is a typical non-orientable surface. This term comes from the integral of (\ref{N0U}), which does
not exist in the unitary case.

We obtain  the connected part of the two-point correlation, 
after the shift $u_i\rightarrow u_i+ \frac{s_i}{2}$ in (\ref{UUU}),
\ba\label{Us1s2}
&&\lim\limits_{N\rightarrow 0}
\tilde U(s_1,s_2) =  - e^{\frac{1}{4}(s_1^2+s_2^2)}\oint
 \frac{du_1 du_2}{2\pi i} e^{ s_1 u_1+s_2 u_2}  \nonumber\\
&\times&   {\rm log}(1+ \frac{ s_1}{2u_1})
\frac{(u_1+\frac{s_1}{2})(u_2+\frac{s_2}{2})}{[u_1^2-(u_2+\frac{s_2}{2})^2]
[u_2^2 -(u_1+\frac{s_1}{2})^2]}
\ea
with
\be\label{usum}
U(s_1,s_2) = \sum_{\epsilon_1,\epsilon_2=\pm 1} \tilde U(\epsilon_1 s_1,\epsilon_2 s_2).
\ee
We deform the contour of $u_2$ near the origin to the poles (i) $u_2=u_1+\frac{s_1}{2}$, (ii)$u_2=-u_1-\frac{s_1}{2}$,
(iii)$u_2=-u_1-\frac{s_2}{2}$,(iv)$u_2=u_1-\frac{s_2}{2}$.
Then we obtain in the replica limit $N\rightarrow 0$,
\ba
\lim\limits_{N\rightarrow 0} \tilde U(s_1,s_2) &=& \frac{2}{(s_1+s_2)^2}{\rm sh} \frac{s_1}{4}(s_1+s_2){\rm sh} \frac{s_2}{4}(s_1+s_2)\nonumber\\
&-& \frac{2}{(s_1-s_2)^2}{\rm sh} \frac{s_1}{4}(s_1-s_2){\rm sh} \frac{s_2}{4}(s_1-s_2)\nonumber\\
&+& \frac{1}{2}\int_{s_1-s_2}^{s_1+s_2}dy \frac{1}{y}{\rm sh}\frac{s_1 y}{4}{\rm sh}\frac{s_2 y}{4}.
\ea
Since this is invariant under the change of signs of $s_i$, we have from (\ref{usum}),
\be
\lim\limits_{N\rightarrow 0} U(s_1,s_2) = 4 \lim\limits_{N\rightarrow 0} \tilde U(s_1,s_2).
\ee
For $n>2$ we operate in a similar fashion and  obtain the following result which generalizes 
the theorem for the unitary case \cite{BH2} : 

{\bf Theorem 3:}
\ba
&&\lim\limits_{N\rightarrow 0} U(s_1,...,s_n) = \sum_{\epsilon_i=\pm 1} 
W(\epsilon_1 s_1,\epsilon_2 s_2,...,\epsilon_n s_n),\nonumber\\
&&W(s_1,...,s_n) = \frac{1}{2\sigma^2}\prod_{i=1}^n 
( 4 {\rm sh}\frac{s_i \sigma}{4})+ \frac{1}{2}\int_0^\sigma dy
\frac{1}{y}\prod_{i=1}^n {\rm sh}\frac{s_i y}{4}
\ea
where $\sigma= s_1+\cdots + s_n$.

%%%%%%%%%%%%%%%%%%%%%%%%%%%%%%%%%%%%%%%%%%%%%%%%%%%%%%%%%%%%%%
\section{Intersection numbers from $U(s_1,...,s_n)$}
\setcounter{equation}{0}
\renewcommand{\theequation}{6.\arabic{equation}}

For  one marked point, we consider the evolution operator $U(s)$ in an external source $A$, chosen at  a critical
value. We discuss here the case p=3.
From (\ref{U(s)}), by the scalings of $v\rightarrow \sqrt{N}v, s\rightarrow s/\sqrt{N}$, and 
by the critical value $a_i^2=N$, we have
\be
U(s) = -\frac{1}{s} e^{\frac{s^2}{4N}}\oint \frac{dv}{2\pi i} 
\left(\frac{1 + v^2}{1 + (v+ \frac{s}{2N})^2}\right)^{N}
\frac{v+ \frac{s}{2N}}{v+ \frac{s}{4N}} e^{sv}
\ee
Exponentiating the term of power $N$, we have
\ba
&&- N {\rm log} [1 + (v+ \frac{s}{2N})^2] +  N {\rm log}(1 + v^2)+ s v + \frac{s^2}{4N}\nonumber\\
&&= v^3 s + \frac{3}{4N}v^2 s^2 + \frac{1}{4 N^2} v s^3 + \frac{1}{32 N^3} s^4 + \cdots
\ea
The first four terms are of order $N$ after rescaling  $s\rightarrow N s$. Further, by the replacement by
$s\rightarrow \sqrt{2}s$,$N\rightarrow 2N$,$v\rightarrow u/\sqrt{2} $, we obtain in the large N limit,
\be
U(s) = -\frac{e^{\frac{N}{4}s^4}}{Ns} \int \frac{du}{2\pi i} 
e^{ N s u^3 + \frac{3}{2}N s^2 u^2 + N s^3 u}\left(
\frac{u+s}{u+\frac{s}{2}}\right)
\ee
The shift $u\rightarrow u-\frac{s}{2}$, and $u=  \frac{1}{s^{1/3}}t$ gives
\be
U(s) = -\frac{1}{N s^{4/3}}\int \frac{dt}{2\pi i}
 e^{Nt^3 + \frac{N}{4}s^{8/3}t}( 1 + \frac{s^{4/3}}{2t}).
\ee
We further make a scale $t\rightarrow - i t/(3N)^{1/3}$, then we have
\ba
U(s) &=&  \frac{1}{3^{1/3}(N s)^{4/3}}[ \frac{1}{\pi}\int_0^\infty dt
 {\rm cos}(\frac{t^3}{3} + x t)\nonumber\\
&&
-\frac{1}{\pi} \frac{s^{4/3}(3N)^{1/3}}{2} \int_0^\infty
dt \frac{1}{t}{\rm sin}(\frac{t^3}{3} + x t)].
\ea
where $x = - N^{2/3}s^{8/3}/(4\cdot 3^{1/3})$.
Using the Airy function $A_i(x)$,
\be
A_i(x) = \frac{1}{\pi}\int_0^\infty dt {\rm cos}(\frac{t^3}{3} + x t)
\ee
we obtain
\be\label{final}
U(s) =  \frac{1}{3^{1/3}(N s)^{4/3}} 
\left( A_i(x) - \frac{s^{4/3}(3 N)^{1/3}}{2}\int_0^x dx^\prime
 A_i(x^\prime)\right)
\ee

Since $s$ is a Fourier transform  variable, it is proportional to
\be
s\sim \frac{1}{\lambda^3}, \hskip 5mm x \sim s^{8/3} \sim \frac{1}{\lambda^8}
\ee
The Airy function $A_i(x)$ has asymptotic expansion
\ba\label{Ai}
A_i(x) &&= A_i(0)(1+ \frac{1}{3!}x^3+\frac{1\cdot 4}{6!}x^6+\cdots + \frac{1\cdot 4\cdot 7}{9!} x^9 + \cdots)
\nonumber\\
&& +
A_i^\prime(0)(x+ \frac{2}{4!}x^4+ \frac{2\cdot 5}{7!}x^7+ \frac{2\cdot 5\cdot 8}{10!} x^{10}+ \cdots)
\ea
where $A_i(0)= 3^{-2/3}/\Gamma(2/3)$ and $A_i^\prime(0)=-3^{-1/3}/\Gamma(1/3)$.

Therefore, we obtain the intersection numbers from the coefficients of the evolution operator $U(s)$ 
for the one marked point like in  the previous unitary case \cite{BH2,BH3}.
The first series of expansion of $A_i(x)$ in (\ref{Ai})gives the intersection 
numbers for the spin $j=1$, and the second
series gives the intersection numbers for the spin $j=0$. 
From the first term in (\ref{final}) with (\ref{Ai}), one finds the intersection numbers $<\tau_{n,j}>_g$  for integer $g$ 
\be
< \tau_{(8 g -5 - j)/3,j} >_g = \frac{1}{(24)^g g!} \frac{\Gamma(\frac{g + 1}{3})}{\Gamma(\frac{2 - j}{3})}
\ee
where $j=0$ for $g= 3 l+1$ and $j=1$ for $g= 3 l$.   The intersection numbers for non orientable surfaces
of half-integer genera, are obtained from the second term of  
(\ref{final}), namely the integral of the Airy function. Taking into account  the normalization (\ref{half}),  this leads to explicit results, such as \\
$$<\tau_{2,1}>_{g=3/2} = \frac{1}{864}$$.
\vskip 5mm
\section {Conclusion}

We have derived the intersection numbers for non-orientable surfaces from generalized 
Kontsevich Airy integrals over  random antisymmetric matrices, 
the Lie algebra of the group $SO(2N)$. An $N-k$ duality bewteen  
k-point functions in $N\times N$ Gaussian matrix integrals, 
and N-point functions for $k\times k$integrals, 
in the presence of an external matrix source, allows one 
to relate those generalized Airy integrals to the edge 
behavior of Gaussian models. Those Gaussian models are then 
much easier to deal with than the original integrals. 
The existence for Lie algebras of classical groups  
(such as Hermitian matrices for U(N) or antisymmetric matrices for O(N))
 of an Harish Chandra integral over the group elements is a key ingredient 
in these calculations. These techniques should be useful for characterizing 
the geometric properties of those non orientable surfaces, and for comparing 
zeros of analytic functions to random matrix spectra. \\ \vskip 3mm
{\bf Acknowledgements}
S.H. thanks  the supports of the Ecole Normale Sup\'erieure and  of 
the Grant-in-Aid for Scientific Research (C19540395) by  JSPS.

%%%%%%%%%%%%%%%%%%%%%%%%%%%%%%%%%%%%%%%%%%%%%%%%%%%%%%%%%%%%%%%%%%%%%%%%%%
\vskip 5mm
%%%%%%%%%%%%%%%%%%%%%%%%%%%%%%%%%%%%%%%%%%%%%%%%%%%%%%%%%%%%%%%%%%%%%%%%%%%%%%%
\vskip 5mm
{\bf Appendix: derivation of theorem 1.}
\setcounter{equation}{0}
\renewcommand{\theequation}{A.\arabic{equation}}
\vskip 3mm

Let us begin with  the simple case, k=1.  The determinant is given by the integral,
\be\label{a2}
<{\rm det}(\lambda\cdot{\rm I} - X)> = <\int dc_a d\bar c_a e^{\bar c_a(\lambda\cdot {\rm I} - X)_{ab} c_b}>
\ee
over the $2N$ Grassmann variables $c_a, \bar c_a$. 
The probability measure $P(X)$ for the average is
\be
P(X) = \frac{1}{Z_A} e^{\frac{1}{2}{\tr} X^2+ {\tr X A}}
\ee
Absorbing the antisymmetric part of the term $\bar c_a X_{ab}c_b$  in  the external source, 
$A \to A^{\prime}$,  with 
\be
A^{\prime}_{ab} = A_{ab} - \frac{1}{2}(\bar c_a c_b - \bar c_b c_a)
\ee
We obtain
\be
{\tr} {A^{\prime}}^2 = {\tr A^2} - 2 A_{ab} \bar c_b c_a - \frac{1}{2}(\bar c_a c_a)(\bar c_b c_b)
\ee
where $a$ and $b$ run over $1,2,\cdots, N$.
Writing the term  $(\bar c_a c_a)(\bar c_b c_b)$ in the exponent as
\be
e^{-\frac{1}{4}(\bar c_a c_a)(\bar c_b c_b)}
= \frac{1}{\sqrt{\pi}}\int dy e^{-y^2 + i y \bar c c}
\ee
we obtain 
\ba \label {lhs}
<{\rm det}(\lambda\cdot {\rm I} - X)> =&&\frac{1}{\sqrt{\pi}}\int dy \int d c d \bar c e^{- y^2 + (\lambda + i y) \bar c_a c_a - A_{ab}
\bar c_b c_a}\nonumber\\
&=& \frac{1}{\sqrt{\pi}}\int dy e^{- y^2} \prod_{j=1}^N {\rm det}
((\lambda + i y) \cdot {\rm 1} - \frac{a_j}{\gamma}i \sigma_2)\nonumber\\
&=&\frac{1}{\sqrt{\pi}} \int dy e^{-y^2} \prod_{j=1}^N [(\lambda + i y)^2 - {a_j^2}]
\ea
where $\sigma_2$ is the second  Pauli matrix.
On the other hand for the one-point function the right-hand side in (\ref{eq1}) (theorem 1)  is an integral over one single real number 
\ba\label {rhs}
<\prod_{j=1}^N {\rm det}(a_j\cdot {\rm I} - Y)>_\Lambda 
&=&\frac{1}{Z_{\Lambda}} \int \prod_{j=1}^N {\rm det} \left( \matrix{ 
a_j & y \cr
-y & a_j }\right)e^{- y^2 - 2 i \lambda y}\nonumber\\
&=&\frac{1}{\sqrt{\pi}} e^{-\la^2} \int dy \prod_{j=1}^N (a_j^2 + y^2) e^{- y^2 - 2 i  \lambda y}.
\ea
The shift  $iy \rightarrow iy +\la$ shows that (\ref{lhs}) and (\ref{rhs}) are identical.

When  $k\ge 2$ for averaging the $k$ characteristic polynomials, 
$k\times 2N$  Grassmann variables $\bar c^\alpha_a$ and $c^\alpha_b$ are necessary
($\alpha = 1,...,k$  ; $a,b= 1\cdots N$). 
As for $k=1$, we have
\be
A^{\prime}_{ab} = A_{ab} - \frac{1}{2} (\bar c^\alpha_a c^\alpha_b - \bar c^\alpha_b c^\alpha_a)
\ee
\be
{\tr} {A^{\prime}}^2 = - 2 A_{ba}\bar c^\alpha_a c^\alpha_b - \frac{1}{2}\bar c^\alpha_a c^\beta_a \bar c^\beta_b c^\alpha_b
- \frac{1}{2}\bar c^\alpha_a \bar c^\beta_a  c^\beta_b c^\alpha_b
\ee
The last two terms are replaced  by the following integrals,
\be
e^{-\frac{1}{4\gamma} (\bar c^\alpha_a c^\beta_a)(\bar c^\beta_b c^\alpha_b)} = \int dB
e^{-\gamma {\tr}B^2 + i {\tr}B^{\beta \alpha}\bar c^\alpha_a c^\beta_a}
\ee
\be
e^{-\frac{1}{4\gamma} (\bar c^\alpha_a \bar c^\beta_a)(c^\beta_b c^\alpha_b)} = 
\int d D d D^* e^{\gamma {\tr} D^* D + \frac{1}{2}\bar c^\alpha_a \bar c^\beta_a D^{\beta \alpha}
+ \frac{1}{2}(D^*)^{\alpha \beta}
c^\beta_b c^\alpha_b}
\ee
where $B$ is a $k\times k$ Hermitian matrix and $D$ is a $k\times k$  antisymmetric complex matrix.
Thus we obtain,
\ba
&&< \prod_{\alpha=1}^k {\rm det}(\lambda_\alpha - X) > = \int dB dD e^{-\gamma {\tr} B^2+ \gamma {\tr} D^*D
- \frac{1}{\gamma} A_{ba}\bar c^\alpha_b c^\alpha_a}
\nonumber\\
&&\times e^{\lambda_\alpha \bar c^\alpha_a c^\alpha_a + i B^{\alpha \beta}\bar c^\beta_a c^\alpha_a
+ \frac{1}{2}D_{\alpha \beta} \bar c^\beta_a \bar c^\alpha_a + \frac{1}{2}D^*_{\alpha \beta}c^\beta_a
c^\alpha_a}
\ea
The exponent of this integral is a quadratic form in the Grassmann variables.

Let us  first consider the $k=2$ case. The exponent is of the form    $\sum_n\Psi_n^t M_n \Psi_n$, where
\be
\Psi_n^t = (\bar c^1_{2n+1}, \bar c^1_{2n+2}, \bar c^2_{2n+1}, \bar c^{2}_{2n+2}, c^1_{2n+1},
c^1_{2n+2}, c^2_{2n+1},
c^2_{2n+2})
\ee
with  the 8 by 8 matrix $M_n$ 
\be
M_n = \left(\matrix{\tilde D& \tilde B_n\cr
    -\tilde B_n^t& \tilde D^*}\right)
\ee
where
\be
\tilde D = \left(\matrix{0&0&D_{21}&0\cr
                        0&0&0&D_{21}\cr
                        -D_{21}&0&0&0\cr
                        0&-D_{21}&0&0}\right)
\ee
\be
\tilde B_n =\left(\matrix{\lambda_1+ i B_{11}& \frac{i}{\gamma} a_n& i B_{21}&0\cr
                      -\frac{i}{\gamma}a_n& \lambda_1+i B_{11}&0&i B_{21}\cr
                        iB_{12}&0&\lambda_2+iB_{22}&\frac{1}{\gamma} a_n\cr
                       0&iB_{12}&-\frac{i}{\gamma}a_n&\lambda_2+ i B_{22}}\right)
\ee

Since the matrix $M_n$ is antisymmetric, the Gaussian integral over the $\Psi_n$ is  the Paffian : 
\ba
{\rm Pf}( M_n ) &=& {\Big[} |D_{21}|^2-(\lambda_1+iB_{11})(\lambda_2 + i B_{22})- |B_{12}|^2{\Big]}^2\nonumber\\
      &&+ \frac{1}{\gamma^2}a_n^2 {\Big[}- (\lambda_1+ i B_{11})^2 - (\lambda_2 + 
i B_{22})^2 + 2 |B_{12}|^2 + 
2 |D_{21}|^2{\Big]}\nonumber\\
&&+ \frac{1}{\gamma^4}a_n^4
\ea
Writing $B_{11}= b_1,B_{22}=b_2,B_{12} = b_3+ i b_4,D=d_1+ i d_2$, 
we define the real antisymmetric matrix $Y$ as
\be
Y = \left(\matrix{0&b_1&b_4+d_2&b_3+d_1\cr
                 -b_1&0&d_1-b_3&b_4-d_2\cr
                 -b_4-d_2&-d_1+b_3&0&b_2\cr
                 -b_3-d_1&-b_4+d_2&-b_2&0}\right)
\ee
which satisfies the following equation,
\be
{\rm det} (\frac{a_n}{\gamma}\cdot{\rm I} - Y) 
= {\rm Pf}( M_n )
\ee
Since 
\ba
{\tr} Y^2 &=& - 2 (b_1^2+b_2^2) - 4 (b_3^2+b_4^2+d_1^2+d_2^2)\nonumber\\
&=& -{\tr }B^2 - {\tr} D^* D
\ea
theorem 1 holds.\\ For $k>2$,  the same procedure  leads to the expression of an antisymmetric matrix Y,
which is made of block of $ 2\times 2$ matrices, given by
\be
Y_{ij} = ({\rm Im}B_{ij})\cdot {\rm I} + ( i \sigma_1){\rm Re}B_{ij} + \sigma_1
{\rm Im}D_{ij} + \sigma_3 {\rm Re} D_{ij}
\ee
where the $\sigma_i$ are the  $2\times2$ Pauli matrices.

\vskip 8mm
{\bf Appendix B: HarishChandra integral formula of theorem 2}
\setcounter{equation}{0}
\renewcommand{\theequation}{B.\arabic{equation}}
\vskip 5mm

The Weyl group $W$ of the $SO(2N)$ Lie algebra  is the permutation group $S_{2N}$ followed by reflection symmetries $ y_i\to \epsilon_i y_i$ ($\epsilon_i =\pm 1$), with an even number of $\epsilon_i=-1$ . Then the sum over the elements of the Weyl group contained in the numerator of (\ref{Th2}) becomes
\be
I =\sum\limits_{{\epsilon_1=\pm 1,\cdots,\epsilon_{N}=\pm 1}
\atop{\epsilon_1 \epsilon_2 \cdots \epsilon_{N} = 1}} \sum_{\sigma\in S_{N}} ({\rm det}\sigma)
{\rm exp}[ 2 \sum_{j=1}^{N} \epsilon_{\sigma(j)}y_{\sigma(j)} \lambda_j] 
\ee
where the sum  is restricted to reflections with an even number of sign changes, i.e.
\be\label{cond}
\epsilon_1 \epsilon_2 \cdots \epsilon_{N} = 1.
\ee
Since ${\rm det}\sigma=(-1)^{\vert \sigma \vert}$ (in which $\vert \sigma \vert$ is 
the parity of the permutation),  the sum over the $N!$ elements $\sigma$  of 
$S_{N}$ is a determinant. Therefore we obtain
\be \label{simple}
I = \sum_{{\epsilon_1=\pm 1,\cdots,\epsilon_{N}=\pm 1}\atop{\epsilon_1 \epsilon_2 \cdots 
\epsilon_{N} = 1}}
{\rm det} [ e^{2 \epsilon_i y_i \lambda_j} ]
\ee
The result in this form is sufficient for the purposes of section 4. But one may go a bit further. 
Writing for each matrix element $e^{2 \epsilon_i y_i \lambda_j}  = \cosh{(2y_i\lambda_j)} + \epsilon_i \sinh{(2y_i\lambda_j) } $ we obtain a sum of $2^{N}$ determinants weighted by products of $\epsilon_i$. The sum over those $\epsilon_i$, restricted by the condition  (\ref{cond}), leads to a cancellation of all the terms except two . The final result is
\be{\label {determinant}}
I = 2^{N-1} ({\rm det}[{\rm cosh}(2 y_i \lambda_j)] + {\rm det}[{\rm sinh}(2 y_i \lambda_j)])
\ee
For instance, for $N=2$
the signs are $(\epsilon_1,\epsilon_2) = (1,1)$ or $(-1,-1)$ . The sum over these
terms gives
\ba
I&=& {\rm det}[e^{2 y_i\lambda_j}] + {\rm det}[e^{-2 y_i \lambda_j}]\nonumber\\
&=& (e^{2 y_1 \lambda_1+ 2y_2\lambda_2}-e^{2 y_1\lambda_2+2y_2\lambda_1})+
(e^{-2 y_1\lambda_1-2 y_2\lambda_2}-e^{-2 y_1\lambda_2 - 2 y_2\lambda_1}).
\ea
which is indeed identical to
\be
I =2( {\rm det}[{\rm cosh}(2 y_i\lambda_j)] + {\rm det}[{\rm sinh}(2 y_i \lambda_j)]).
\ee
\vskip 5mm

Let us apply the result (\ref{simple}) to the integral (\ref{Z_p}) :
\be
Z= \int dY e^{-\frac{1}{p+1}{\tr} Y^{p+1} + {\tr} Y \Lambda}
\ee
in which $Y$ runs over the $2N\times 2N$ antisymmetric matrices, the Lie algebra of $SO(2N)$. 
One may use the rotational invariance of the measure to write $Y = g y g^{-1}$ in which $g$ is 
an element of $SO(2N)$ and  $y$ is a canonical matrix (\ref {canonical}), namely 
$$y = y_1 v \oplus  \cdots \oplus y_N v, \hskip 5mm v =i\sigma_2 = \left( \matrix{0&1\cr -1&0}\right ).
$$
The integral over $Y$ may be replaced by an integral over $g$ and over the $y_i$'s. 
The Jacobian is, up to a constant factor, $ J= \prod_{i<j}(y_i^2-y_j^2)^2$. Using the 
Harish Chandra integral, one integrates over $g$  and, using the result (\ref{simple}), 
one obtains
\be
Z= \sum_{{\epsilon_1=\pm 1,\cdots,\epsilon_{N}=\pm 1}
\atop{\epsilon_1 \epsilon_2 \cdots \epsilon_{N} = 1} }\int dy_1
\cdots dy_N \frac{\prod(y_i^2-y_j^2)}{\prod (\lambda_i^2-\lambda_j^2)} 
e^{-\frac{2}{p+1}\sum_1^N y_i^{p+1}  } {\rm det} [ e^{2 \epsilon_i y_i \lambda_j} ].
\ee
Since $p$ is odd, one can change $\epsilon_i y_i \to y_i$. 
The antisymmetry of ${\prod(y_i^2-y_j^2)}$ under permutations 
allows one to replace the determinant by its diagonal term 
$\rm {exp}(2\sum y_i\lambda_i)$ (up to a factor $N!$).
 The restricted sum over the $\epsilon_i$ is simply the factor $2^{N-1}$.
 We are thus led to the integral (\ref{postHC}) of section 4. 
\vskip 5mm
{\bf Appendix C: integral representations for $U(s_1,\cdots,s_n)$}
\setcounter{equation}{0}
\renewcommand{\theequation}{C.\arabic{equation}}
\vskip 5mm

For real antisymmetric matrices, the support of the density of states $\rho(\lambda)$ is the imaginary axis,
\be
\rho(\lambda) = \frac{1}{N}< {\rm tr} \delta(\lambda - X) >.
\ee
and it is an even function of $\lambda$. 
The Fourier transform of $\rho(\lambda)$ is $U(t)$, 
\ba
U(t) &=& \int d\lambda e^{i t \lambda} \rho(\lambda)\nonumber\\
&=& \frac{1}{N} < {\rm tr} e^{i t X} >.
\ea
Through an orthogonal transformation $g$ $(g\in SO(2N))$, one can bring the matrix $X$ to the canonical form
\be
X = x_1 v \oplus  \cdots \oplus x_N v, \hskip 5mm v =i\sigma_2 = \left( \matrix{0&1\cr -1&0}\right ).
\ee
 from which one has
\be
{\rm tr} e^{s X} = 2 \sum_{i=1}^N {\rm cos}(s x_i)
\ee
From  theorem 2, the evolution operator $U(s)$ becomes
\be
U(s) = \frac{1}{N}\sum_{\alpha=1}^N \int \prod_i dx_i {\rm cos}(s x_\alpha) \frac{\Delta(x_j^2)}{\Delta(a_j^2)}
e^{- \sum x_i^2 + 2 \sum a_j x_j }
\ee
If we write ${\rm cos}(sx_{\alpha}) = {\rm Re}  ( e^{i s x_{\alpha}})$ the integral amounts to an integral over antisymmetric matrices with the shifted source $a_j\to \tilde a_j =  a_j+\frac{1}{2} is\delta_{j,\alpha}$. Then the  integrals over the $x_i$'s  are just the normalization for Gaussian antisymmetric matrices in the source $\tilde A$  :
\be \int dX e^{ \frac{1}{2} {\rm Tr} X^2 + {\rm Tr}\tilde AX} = e^{-\frac{1}{2} {\rm Tr} \tilde A^2} .\ee
Indeed using again the Harish Chandra theorem, the left-hand side is simply
\be  \int dX e^{ \frac{1}{2} {\rm Tr} X^2 + {\rm Tr}\tilde AX} = \int \prod_i dx_i  \frac{\Delta(x_j^2)}{\Delta(\tilde a_j^2)}
e^{- \sum x_i^2 + 2 \sum \tilde a_j x_j }\ee
which provides the result that we need :
\be \int \prod_i dx_i  \Delta(x_j^2)
e^{- \sum x_i^2 + 2 \sum a_j x_j +isx_{\alpha}} =  \frac{\Delta(\tilde a_j^2)}{\Delta(a_j^2)} e^{-\frac{1}{2} {\rm Tr} \tilde A^2 +\frac{1}{2} {\rm Tr} A^2} .
\ee
This leads to 
\be\label{uus}
U(s) = \frac{1}{2N} \sum_{\alpha=1}^N \prod_{\gamma \ne \alpha}^N 
\left( \frac{(a_\alpha + \frac{i s}{2})^2- a_\gamma^2}{
a_\alpha^2-a_\gamma^2}\right) e^{i s a_\alpha - \frac{s^2}{4}} + (s\to -s).
\ee
The normalization is such that $U(0)=1$. It is useful to express (\ref{uus}) as a  contour integral,
\be\label{C10}
U(s) = \frac{1}{N s} \oint \frac{du}{2\pi i} \prod_{\gamma=1}^N \left( 
\frac{(u+\frac{i s}{2})^2 - a_\gamma^2}{
u^2 - a_\gamma^2} \right) \frac{u}{i u - \frac{s}{4}} e^{i u s - \frac{s^2}{4}}
\ee
where the contour encircles the poles $u = a_\gamma$.

Changing  variables, $u + \frac{i s}{2} = - i v$, one obtains  $U(s)$
\be\label{C1}
U(s) = -\frac{1}{N s} \oint \frac{dv}{2\pi i}\prod_{n=1}^N \left( \frac{v^2+ a_n^2}{
(v+ \frac{s}{2})^2 + a_n^2}\right)
\frac{v + \frac{s}{2}}{v+ \frac{s}{4}}e^{v s + \frac{s^2}{4}},
\ee
which is the representation that we have used in (\ref{U(s)}).

The two-point correlation function $U(s_1,s_2)$ is given by Th.2,
\ba\label{UU}
&&U(s_1,s_2)= \frac{1}{2N}<{\rm tr} e^{s_1 X} {\rm tr} e^{s_2 X} >\nonumber\\
&=& \frac{2}{N}\sum_{\alpha_1,\alpha_2=1}^N \int \prod_{i=1}^N dx_i {\rm cos}(s_1 x_{\alpha_1})
{\rm cos}(s_2 x_{\alpha_2})\frac{\Delta(x^2)}{\Delta(a^2)}e^{- \sum x_i^2 + 2 \sum a_j x_j}.\nonumber\\
\ea
We make  replacements of ${\rm cos}(i s_i x_{\alpha_i})$ by 
$ \frac{1}{2}e^{i s_i x_{\alpha_i}}, (i=1,2)$
in (\ref{UU}), and we name it as $\tilde U(s_1,s_2)$. Then, we have
\be
U(s_1,s_2) = \sum_{\epsilon_1,\epsilon_2=\pm 1} \tilde U(\epsilon_1 s_1,\epsilon_2 s_2).
\ee
The sum in (\ref{UU}) is devided into two parts, $\sum_{\alpha_1=\alpha_2}$ and $\sum_{\alpha_1\ne \alpha_2}$.
The first part can be neglected.
By the double contour integrals, the double sum is expressed as
\ba
&&\tilde U(s_1,s_2) = \frac{1}{2 N s_1 s_2}\oint \frac{du dv}{(2\pi i)^2}\prod_{\gamma=1}^N
\left( \frac{(u+\frac{i s_1}{2})^2-a_\gamma^2}{u^2-a_\gamma^2}\right)\left(
\frac{(v+ \frac{i s_2}{2})^2 -a_\gamma^2}{v^2-a_\gamma^2}\right)\nonumber\\
&\times&\frac{u v}{(i u - \frac{s_1}{4})(i v - \frac{s_2}{4})}
 \frac{[(u+\frac{i s_1}{2})^2- (v + \frac{i s_2}{2})^2](u^2-v^2)}{
[(u+\frac{i s_1}{2})^2 - v^2][u^2-(v+\frac{i s_2}{2})^2]}
e^{i u s_1 + i v s_2 - \frac{1}{4}(s_1^2+s_2^2)}\nonumber\\
\ea
By the Cauchy determinant identity,
\be
{\rm det}\frac{1}{x_i^2-y_j^2} = (-1)^{n(n-1)/2}\frac{\prod_{i<j}(x_i^2-x_j^2)(y_i^2-y_j^2)}{
\prod_{i,j}(x_i^2-y_j^2)}
\ee
with $x_i=u_i+ \frac{i s_i}{2}, y_i = u_i$, above expression is simplified as
\ba
&&\tilde U(s_1,s_2) = \frac{1}{2 N} e^{-\frac{1}{4}(s_1^2+s_2^2)}\oint
\frac{du_i}{(2\pi i)^2} e^{\sum i s_i u_i}  \prod_{i=1}^2 u_i\nonumber\\
&&\times \prod_{\gamma=1}^N \prod_{i=1}^2 \left( \frac{(u_i+ \frac{i s_i}{2})^2-a_\gamma^2}{
u_i^2 -a_\gamma^2}\right)
{\rm det}\frac{1}{(u_i + \frac{i s_i}{2})^2 - u_j^2}
\ea
For general n, we have similarly,
\ba\label{C16}
&&\tilde U(s_1,\cdots,s_n) = \frac{1}{ 2 N} e^{-\frac{1}{4}\sum s_i^2}\oint
\prod_{i=1}^N \frac{du_i}{2\pi i} e^{\sum i s_i u_i}  \prod_{i=1}^n u_i\nonumber\\
&&\times \prod_{\gamma=1}^N \prod_{i=1}^n \left( \frac{(u_i+ \frac{i s_i}{2})^2-a_\gamma^2}{
u_i^2 -a_\gamma^2}\right)
{\rm det}\frac{1}{(u_i + \frac{i s_i}{2})^2 - u_j^2}.
\ea
with
\be
U(s_1,\cdots,s_n) = \sum_{\epsilon_i=\pm 1} \tilde U(\epsilon_1 s_1,\cdots,\epsilon_n s_n).
\ee
This expression reduces to  the previous one in (\ref{C10}) for n=1.
We need the connected part of $U(s_1,\cdots,s_n)$, which is easily obtained from the expression
of the determinant
in (\ref{C16}). For n=2, we obtain by the change $u_i\rightarrow - i u_i$,
\ba\label{UUU}
&&\tilde U_c(s_1,s_2) = \frac{1}{2 N} e^{-\frac{1}{4}(s_1^2+s_2^2)}\oint
\frac{du_i}{(2\pi i)^2} e^{\sum  s_i u_i} \nonumber\\
&&\times \prod_{\gamma=1}^N \prod_{i=1}^2 \left( \frac{(u_i- \frac{ s_i}{2})^2 + a_\gamma^2}{
u_i^2 + a_\gamma^2}\right)
\frac{u_1 u_2}{[(u_1-\frac{s_1}{2})^2-u_2^2][(u_2-\frac{s_2}{2})^2-u_1^2]}
\ea
This is the representation that we have used in (\ref{Us1s2}).

\vskip 5mm
%%%%%%%%%%%%%%%%%%%%%%%%%%%%%%%%%%%%%%%%%%%%%%%%%%%%%%%%%%%%%%%%%

\end{document}